# Energy resolution and energy-light response of CsI(Tl) scintillators for charged particle detection

A. Wagner[a1], W.P. Tan[1], K. Chalut[1], R.J. Charity[3], B. Davin[2], Y. Larochelle[c2], M.D. Lennek[1], T.X. Liu[1], X.D. Liu[1], W.G. Lynch[1], A.M. Ramos[1], R. Shomin[1], L.G. Sobotka[3], R.T. de Souza[2], M.B. Tsang[1], G. Verde[1], H.S. Xu[1]

[1] *National Superconducting Cyclotron Laboratory and Department of Physics and Astronomy, Michigan State University, East Lansing, MI 48824, USA.*
[2] *Department of Chemistry and IUCF, Indiana University, Bloomington, IN 47405, USA.*
[3] *Department of Chemistry, Washington University, St. Louis, MO 63130, USA.*

**Abstract:**

This article describes the crystal selection and quality control utilized to develop and calibrate a high resolution array of CsI(Tl) scintillator crystals for the detection of energetic charged particles. Alpha sources are used to test the light output variation due to thallium doping gradients. Selection of crystals with better than 1% non-uniformity in light output is accomplished using this method. Tests with a 240 MeV alpha beam reveal that local light output variations within each of the tested CsI(Tl) crystals limit the resolution to about 0.5%. Charge and mass dependences in the energy - light output relationship are determined by calibrating with energetic projectile fragmentation beams.

---

[a] Present address: Institut für Kern- und Hadronenphysik, Forschungszentrum Rossendorf, D-01314 Dresden, Germany
[c] Present address: Laboratoire de Physique Nucléaire, Université Laval, Canada, G1K-7P4.

**Introduction:**

CsI(Tl) scintillation detectors are a cost effective technology for detecting charged particles with energies of E/A=30-200 MeV[1-11]. Less expensive than solid state detectors, less hygroscopic than NaI(Tl) crystals, and easily machined into different shapes, CsI(Tl) crystals have been incorporated in many large solid angle detection arrays [1-11]. In such applications, greater stability is achieved by avoiding problems related to the temperature dependence [12] of the CsI(Tl) light output by holding the temperature constant, and by reading out the detectors via photodiodes instead of photomultipliers, whose gains may vary with time. In the present article, discussion will be focused upon properties of CsI(Tl) crystals read out by photodiodes.

Energy resolution is an essential requirement of many experiments. For low energy particles, the energy resolutions of crystals, read out by photodiodes, are mainly limited by electronic noise. The importance of this noise depends on the light collection efficiency of the crystal-diode assembly and on the photodiode capacitance. For example, a resolution of 4.4% (59 keV) FWHM was reported [14] for the detection of 1.33 MeV γ rays in a small crystal (3 cm$^3$) with a small (1 cm$^2$) photodiode [14]. Somewhat worse energy resolution 25% (165 keV) FWHM was achieved for the detection of 0.66 MeV γ rays in a much larger 100 cm$^3$ CsI crystal [2] with a larger (2x2 cm$^2$) photodiode [2].

With increasing energy deposition, photon statistics makes an increasing contribution to the resolution that can be approximated by $\delta E = \sqrt{E \cdot E_0}$ where $E$ is particle energy and $E_0$ is the typical energy per photoelectron-hole pair. This latter constant depends on Thalium doping, the light collection of the detector assembly and the quantum efficiency of the photodiode. For reasonable values for $E_0$ of about 70 eV, photon statistics provides a 84 keV contribution to overall energy resolution for particles with $E \approx 100$ MeV. If other factors did not contribute significantly to the resolution, one might expect to achieve resolutions of about several hundred keV in CsI(Tl), which would reduce the incentives to utilize more expensive solid state detector technologies in high resolution experiments.

Unfortunately, the resolutions achieved for higher energy particles are larger than one might expect from noise and photon statistics. For example, energy resolutions of 1.2% (1.2 MeV) FWHM were achieved for 98 MeV α particles with small (1 cm$^3$) crystals [13]. Energy resolutions of 0.8% (740 keV) FWHM were achieved for 92 MeV α particles with larger (100 cm$^3$) CsI crystals [2]. In the latter measurements, the resolutions were comparable to the measured variations in the light output over the volume of the crystals, suggesting that the light output uniformity of the CsI(Tl) scintillator may be a limiting factor in the energy resolution.

As discussed in this article and earlier studies [1,2], reasonably uniform CsI(Tl) crystals can be obtained commercially. Testing and pre-selecting crystals before construction can further improve the overall quality of the crystals. We explore whether this is sufficient to achieve resolutions that are limited by noise and photon statistics and find local non-uniformities in the light output that prevent optimum crystal performance. We also investigate the non-linear mass and charge dependence of the light output of the crystals, which is another factor complicating the use of CsI(Tl) crystals for the detection of charged particles. We find that these dependencies can be constrained by careful energy calibration using beams of different isotopes.

**Preselection of CsI(Tl) crystals for Light Output Uniformity:**

Typically, commercial CsI(Tl) crystals can manifest non-uniformities in the light output across the detector face on the order of one percent per centimeter [1,2]. To a large extent this non-uniformity can be limited to better than 0.3% per centimeter by controlling the manufacturing process and by scanning the CsI(Tl) crystals and rejecting those that do not meet this criterion[1,2]. The influence of this small (<0.3%) residual light output non-uniformity can be determined by combining the CsI(Tl) scintillator with a position sensitive silicon detector. This effect can then be removed by making position dependent corrections to the light output.

The magnitude of the observed non-uniformity is influenced by both the choices of radioactive source and readout scheme [2]. Unlike the energy deposition of an alpha particle which is relatively localized, the energy deposition of a gamma ray samples a larger volume of the crystal and hence the sensitivity to measuring local uniformities is reduced [2]. Photodiodes are more sensitive than phototubes to the stronger light output variations manifested by the longer wavelength scintillation photons [15]. In this section, we describe the pre-selection procedure which involved scanning with a collimated $\alpha$ source and reading out the CsI(Tl) crystal with a silicon photodiode as described in Ref. [2].

As delivered by the manufacturer [16], the crystals were rectangular in shape with dimensions of 3.5 x 3.5 x 6 $cm^3$. They were polished on one 3.5 x 3.5 $cm^2$ surface (here labeled as the front) and sanded at the 3.5 x 6 $cm^2$ sides. Before scanning, the crystals were inspected for visual cracks or imperfections. Then the remaining 3.5 x 3.5 $cm^2$ surface (here labeled as the back) was sanded down and polished. It was then optically coupled to a clear 1x3.5x3.5 $cm^3$ acrylic light guide with optical grease. This light guide was in turn optically connected to a 2 x 2 $cm^2$ photo-diode [17]. The sides of the crystals and the light guide were wrapped with two layers of 0.1 mm thick white Teflon tape. The front face of the crystal was covered with an aluminized mylar foil (0.17 mg/$cm^2$) to ensure uniform light collection.

The 5.486 MeV $\alpha$-particles from a collimated $^{241}$Am $\alpha$-source were used to irradiate the front face of the crystal and monitor the uniformity in the light output of the crystals. The collimators were selected so as to illuminate circular regions of 5mm diameter on the front surface of the detector. The alpha spectra were recorded with a multichannel analyzer equipped with a peak sensing ADC. The spectra were then transferred to a computer and analyzed offline. Figure 1 shows the scanning results of two crystals, #652 that was accepted (top panel) and #291 that was rejected (bottom panel). The peak location of the 5.486 MeV α-line was detected in vacuum at nine equally spaced positions on the crystal face. The scanned position corresponded to the center of each of the nine sub-squares in Figure 1.

The different gray levels of the big squares correspond to the percentage deviations of the alpha peak of each point from the median value. The actual deviations are recorded (in percent) in the small 3x3 table next to the shaded crystal face. The accepted crystal (top panel) is nearly uniform in the shading, varying in light output from –0.11% to 0.05% of the mean. On the other hand, the rejected crystal (bottom panel) clearly shows the existence of a gradient in the light output non-uniformity from left to right. The total variations in light output deviate by nearly ±1% from the mean. Selected crystals were also scanned on the back surface of the crystal. The light output variations displayed by the front and back surfaces were consistent with the component of the thallium doping gradient parallel to the front and back surfaces being approximately uniform throughout the crystal, as observed in Ref. [1,2]. No information was obtained on the doping gradient perpendicular to the front and back surfaces since the influence of that gradient can be addressed by adjustments to the energy- light output relationship.

Crystals with deviations larger than ±0.5% such as the one shown on the bottom panel of Fig. 1 were rejected and sent back to the manufacturer. Crystals with deviations less than ±0.5% were accepted and subsequently machined to their final shapes. This machining step only involves two adjacent sides, which were tapered on an angle of about 7° relative to the normal to the surface of the crystal. This tapering enabled the crystals to form an array of four crystals as shown in Figure 2. Such design allows the crystals to be placed, during subsequent experiment, behind a 5 x 5 cm$^2$ silicon detector forming a ΔE-E telescope that could be closely packed with other telescopes of similar construction at a distance of 20 cm from the target. Consistent with this requirement, the front surface of each CsI(Tl) crystal was reduced to an area of 2.5 x 2.5 cm$^2$ ; the back surface was not modified. The crystals were then polished and scanned one more time. In general, the differences between the results of the initial and final scans were negligible.

**Wrapping materials used for the CsI(Tl) crystals:**

To obtain optimal light collection efficiency for low energy particles [2], a reflective entrance foil was needed on the front face of a CsI(Tl) crystal [2]. Following ref. [2], the sides of the crystal are uniformly sanded with 400-grit wet/dry sand paper using motions parallel to the long axis of the crystal. Several wrapping materials and techniques were tested to see what provided the most suitable diffuse reflecting surface. In one test, the sides of the CsI(Tl) crystals were wrapped, following refs. [1,2], with several layers of 0.1mm thick Teflon tape to minimize light loss and cross-talk between crystals. The upper panel of Fig. 3 shows the signal amplitude (in channels) and the bottom panel shows the resolution (in percent) of 5.486 MeV $\alpha$ particles from a collimated $^{241}$Am $\alpha$-source as a function of the number of layers of Teflon tape used to wrap the crystal. In general, increasing the number of layers of Teflon tape increases the light collection efficiency resulting in larger signal amplitudes (higher peak channels). The improvement saturates at about five layers of Teflon tape. The increase in the percent resolution of the crystal is directly related to the increased light output. For these small signals, the resolution (in channels) is dictated by the electronic noise and is unchanged at about 40 channels (~250 keV) FWHM throughout the test. (This is equivalent to a $\gamma$-ray energy resolution of about 160 keV.)

Five layers of Teflon tape cause a rather thick gap between adjacent CsI(Tl) crystals. Particles impinging on this gap are lost resulting in a loss of detection efficiency. To reduce this gap, we tested the optical properties of 0.14 mm thick cellulose Nitrate membrane with pore size of 0.2 $\mu$m [18]. (Cellulose nitrate achieves its high reflectivity by virtue of the many micropores in the material. Care must be taken not to wet the surfaces of this material or the reflectivity will be reduced.) As shown by the solid points in Fig. 3, significantly improved light collection efficiency and energy resolution are thereby obtained.

For this material, the light output saturated at two layers of cellulose nitrate membranes. Nearly equivalent light output was observed in a wrapping consisting of one layer of cellulose nitrate plus one layer of aluminized mylar. In the final wrapping, each crystal is wrapped with two layers of cellulose nitrate on the outer two surfaces and one layer on the inner surfaces. One layer of aluminized mylar was inserted between adjacent crystals to improve optical isolation. A robust wrapping for 4 crystals assembly is shown in Figure 2.

For the tests with accelerator beams, photodiodes were glued to the light guides with Silicon RTV615. To prevent light leak and cross talk between neighbor crystals, the back face, light guide and the photodiode were painted with a reflective white paint (BC620) from Bicron [19]. The electronic signals from the photodiodes are amplified with charge-sensitive preamplifiers that were connected to the detector using short (6 cm) cables and situated inside the vacuum chamber. The amplified signals are then shaped and amplified by a computer-controlled 16-channel CAMAC shaping amplifier module [20], with a unipolar pulse of 2 μsec shaping time and analyzed by a peak-sensing ADC (Phillips P/S 7184 [21]). The stability of the setup is continuously monitored via a precision pulse generator system and via temperature sensors attached to the detector mounts within the vacuum system.

**<u>Position Dependence of the Energy resolution</u>**

To measure the energy response of the crystals for energetic beams, 240 MeV α particles extracted from the NSCL K1200 cyclotron were injected directly into the CsI(Tl) crystals. Because 1% light output non-uniformities are equivalent to a 2.4 MeV variation in the α particle energies, it was necessary to determine the point of interaction in the detector for each α particle and make corrections to the resulting light output variation. To search for and identify any position dependence in the crystals light output, the position information of each α particle was measured by passing the α's through a 500μm 2 dimensional position sensitive silicon detector (Micron design W [22]) placed in front of the CsI(Tl) crystals.

The double-sided Si-strip detector has 16 strips in the x- and 16 strips in the y-direction. These strips provide 256 co-ordination points (pixels). As the front face of the CsI(Tl) crystal is one-quarter the surface area of the Si-strip detector, 64 measurements of the light output were obtained as a function of position for each crystal.

The right panel of Fig. 4 shows the variations in percent of the light output measured at each of the 64 pixels of crystal #655 for the 240 MeV $^4$He beam. This crystal is chosen for display here since it was illuminated very uniformly by the α beam. Other crystals also display similar trends. This crystal has approximately the same overall uniformity as crystal #652 shown in Fig. 1. Here, however, the sensitivity of the shading levels has been increased enabling the small overall deviations measured in this crystal to be easily observed. The corresponding α source scanning measurements are shown in the left panel. In contrast to the α source measurements that display rather smooth variations, the α beam measurement show significant local variations in the light output, of the order of 0.5%. Some of the local variations are larger than the average change that one observes in the light output from one side of the detector to the other.

This average trend appears to be approximately the same in both α source and α beam measurements. (The dotted lines in the left side for the α source designate the outlines of the front face of the crystal after it was machined to its final shape for the α beam measurements.) To show that these variations are not an experimental artifact and that they are indicative of real variations in the light output of the crystal, the upper and lower panels of Fig. 5 shows energy spectra obtained for pixels along column "*2*" and row "*e*" as labeled in the right panel of Fig. 4, respectively. The exact coordinates of the pixels are labeled inside each panel of the figures. Neighboring pixels correspond to trajectories that are on the average, separated by 3 mm at the front face of the CsI(Tl) crystal. To provide a fixed reference point, the average peak position of the alpha particles detected by the whole crystal is marked by a dashed line (Channel 1315) in each panel. Since different pixels are exposed to different number of particles, the counts in the peak of each pixel are normalized to 1. For clarity, only the statistical error bars of the peaks are indicated. Clearly, there are shifts in these individual spectra, going from one pixel to another, that exceed the resolution of the spectra. Moreover, the trends were not monotonically varying from one direction to another. While the light output near the edge of the crystals (in the extreme left and right panels of Fig. 5) may be sensitive to imperfections in the surface treatment in the crystal, variations in the light output elsewhere must be correlated to local light output variations in the crystal. We speculate that these variations arise from local variations in the thallium doping introduced during the crystal growing. We cannot exclude, however, that these variations could be the result of local impurities that could vary with position on the scale of 3 mm. In either case, one might expect equivalent variations in the light output along the unobserved longitudinal axis of the crystal. Thus one might expect the local variations in the light output to be different for different energy particles, reflecting their different ranges.

The bottom panel of Fig. 6 shows the energy spectra for the 240 MeV $^4$He beam particles detected with CsI crystal #655 without selection on position. The energy resolution is about 1.54 MeV (0.65%). Several attempts can be made to improve this resolution. First, one can correct the light output for the average trend. This was accomplished by fitting the light output with a 3 parameter function $L=L_0(1+ax)(1+bx)$. The spectra are then corrected for each pixel by the relation

$$Ch'= \frac{Ch}{(1+ax)\cdot(1+bx)}. \qquad (1)$$

After making this correction and summing the data from all pixels into one spectrum, there is a slight improvement in the resolution, from 0.65% to 0.59% (1.41 MeV). This corrected spectrum is shown in the middle panel. Alternatively, one can correct the energy-light output pixel by pixel by correcting the energies of particles in one pixel by the ratio of the average energy in that pixel divided by the average energy in the entire crystal. When this was done, the resolution improves dramatically as shown in the top panel, to 0.45% or 1.08 MeV, about twice the noise width of 500 keV. In comparison, the peaks corresponding to the single pixel spectra as shown in Figure 5, have a resolution of 1.04 MeV, nearly the same as the overall resolution obtained after summing up all the single spectra. Since the typical beam energy width is better than 0.1%, the resolution of 0.45% probably represents the upper limit of the resolution of the crystals. Unfortunately, the possible depth dependence of the light output variation excludes the possibility to generalize such corrections to all particles emitted in nuclear reactions.

**Energy calibration:**

The fluorescent light emitted by the CsI(Tl) crystal has two major decay time constants, a fast (~500 ns) and a slow (~7 μs) component. Both components have a light output – energy relationship that is mass and charge dependent. This property has been exploited to provide mass identification for light ions using pulse-shape discrimination [1, 3-11]. The pulse-shape discrimination capability of CsI(Tl) is not needed if one uses the CsI crystals as the stopping detectors in ΔE-E telescopes where Si detectors are used as ΔE detectors. However, the pulse shape dependence on mass remains important because of the influence it has on the energy calibration.

The temporal decay of the CsI(Tl) light output depends on the ionization density, therefore, the charge, mass and energy, of the detected particles [13, 24-29]. At low energy, the light response (L) of a CsI(Tl) crystal is known to show a non- linear correlation with the deposited energy (E), especially for heavy ions, and a dependence of such correlation on both the charge Z and mass A of the detected particle [24]. It also depends on the Tl doping of the crystal.

To determine the energy calibration for different ions, the detectors were directly exposed to low intensity (1000 particles/sec) beams of different isotopes and energies. These ions were obtained by fragmenting 2160 MeV $^{36}$Ar and 960 MeV $^{16}$O primary beams from the NSCL K1200 cyclotron in the A1200 fragment separator [30]. The main advantage of this method is the availability of a large number of particles that could be detected simultaneously (up to 52 isotopes were identified in the case of the $^{36}$Ar fragmentation). Since particles are selected only by their magnetic rigidity (B*ρ= 1.841 Tm for the $^{36}$Ar beam and B*ρ=1.295 Tm for the $^{16}$O beam) one obtains a broad range of different isotopes and energies. The FWHM of the momentum widths for these particles were selected to be 0.5%. The atomic and mass numbers as well as energies of the particles used to calibrate the CsI crystals in the present work are listed in Table I. Hydrogen and helium isotopes were also calibrated by elastic scattering of E/A=30 MeV p-$^4$He molecular beams on a Au target and by 240 MeV direct $^4$He beam particles. The energy calibration for each isotope was done following the mass and charge dependence of the light output described in ref. [24], which in turn was based on previous studies of the light emission of CsI-crystals and on semi-empirical model proposed by Birks [29]. In this approach, the incident particle energy E is parameterized as a function of the light output L, the charge Z, and the mass A of the particle, as

$$E(L,Z,A) = aAZ^2 L + b \cdot (1 + cAZ^2) \cdot L^{1-d \cdot \sqrt{A \cdot Z^2}} \qquad (2)$$

where *a, b, c* and *d* are the fitting parameters with values greater than zero. This expression describes a linear part, dominating at high energies and an exponential part dominating at low energies.

In Fig. 7, the solid and dashed lines represent the best fit of Eq. 2 to the experimental energy calibration data corresponding to different carbon isotopes (A=11-14). The need for a mass dependence can be demonstrated by examining the light output of the higher energy carbon isotopes. At high energy, the light response is expected to be linear. Both the $^{11}$C points should lie in the linear domain. However, a straight line joining the two $^{11}$C isotopes does not pass through the high-energy $^{12}$C, $^{13}$C, and $^{14}$C isotopes. A curve going through all points for the $^{11\text{-}14}$C would lead to a very large and unreasonable curvature compared to calibration procedure adopted elsewhere in the literature. The only solution is a mass dependent calibration curve. Since several fragmentation beams would be required to have the full calibration curve for each isotope, we adopt the mass dependent ansatz (closely related to the quenching effect) of Ref. [24].

For light charged particles with Z≤3, the parameterization described in Eq. 2 did not accurately describe the detected energies. Compared to the observation of Ref. [24], a less pronounced isotopic effect was observed for light ions. This may be the result of the increased concentration of the activator element, Tl, in the CsI-crystals used in the present study compared to those studied in ref. [24]. We find the mass dependence to be over-estimated by the $AZ^2$ factor in Eq. 2, and employ a modified function of Eq. 2 with a weaker dependence on A to fit Z≤3 particles. The expression was modified for each element. For example, for Lithium (Z=3) particles, the first term of Eq. 2 is changed.

$$E(L,Z,A) = a\sqrt{AZ^2}L + b\cdot(1+cAZ^2)\cdot L^{1-d\cdot\sqrt{A\cdot Z^2}} \qquad (3)$$

For Helium (Z=2) isotopes, the following expression was used

$$E = aL + bA^c[1-e^{dL}], \qquad (4)$$

The variables *a, b, c,* and *d* in Eq. 2-4 are fit parameters. There are sufficient data to reproduce with good accuracy the light-output response for all the isotopes of the same element using Equations 2-4. Our fitting procedure resulted in a precision of the energy calibration better than 2% for isotopes from α to O.

As we have only limited calibration points for p, d and t, two calibration points from each isotope, we adopt the simple linear function for Z=1 particles.

$$E = aL + b. \qquad (5)$$

where *a,* and *b* are fit parameters.

More accurate energy calibration of Z=1 particles may be obtained in the future. The present work focuses mainly on heavier elements where a direct calibration with beam fragments is available.

**Summary**

In this article we have described the procedures used to construct CsI(Tl) detectors with good energy resolution. These procedures involve pre-selecting CsI crystals used in the construction of detectors by scanning with an $^{241}$Am alpha source. Global correction factors were thereby obtained which can compensate for the energy resolution due to non-uniformity of light output. In addition, various common wrapping materials are compared in order to obtain a low light cross-talk between adjacent modules with a minimal amount of material between them. The choice of cellulose nitrate micropore filter appears to provide a high reflectivity, much higher than that of white Teflon tape of the same thickness.

Measurements of the resolution with 240 MeV $^4$He beams that can penetrate into the interior of the CsI crystal shows energy resolution of the order of 0.5% can be obtained and suggests that the current CsI crystal manufacturing process produces local variations in the Tl concentrations that are probably depth dependent. Unfortunately, such local variations cannot be corrected easily and they present the major limitations in the energy resolution of the CsI crystals.

Using fragmentation products ranging from hydrogen to oxygen isotopes produced in fragmentation beams and direct alpha and proton beam particles, the relationship between the light response in the CsI crystals and the mass, charge and energy of the detected particles has been investigated. For heavy particles with Z≥3, the mass dependence of the light response function of the CsI crystals cannot be neglected.

The authors wish to thank the accelerator crew at NSCL for providing excellent beams from the K1200 cyclotron and especially Dr. Mathias Steiner for the preparation and delivery of the secondary beams. This work is supported by the National Science Foundation under Grant No. PHY-95-28844 and Department of Energy under Grant No. (DOE-92ER-40714).

**Figure Captions:**

Figure 1: Results of alpha source scans for two crystals. The non-uniformity gradient in Crystal # 652 (top panel) is less than ±0.5% and was accepted. However, the scanning results for Crystal # 291 (bottom panel) is outside ±0.5% non-uniformity and was rejected.

Figure 2: Photograph of an assembled array of four closely packed CsI crystals.

Figure 3: Dependence of light output and resolution of crystals on wrapping materials.

Figure 4: Variations in light output from the scanning of 5.486 MeV alpha particles (left panel) and from 240 MeV $^4$He particles pixel by pixel (right panel). The 8x8 pixel shaded area in the right panel corresponds to the area enclosed by the dashed line in the left panel. Column numbers (*1-8*) and row letters (*a-h*), used to identify each pixel in Figure 5, are marked.

Figure 5: Energy spectra of 240 MeV $^4$He particles for individual pixels. The upper panel shows 8 spectra down column *"2"* and the lower panel shows 8 spectra across Row *"e"*. See Figure 4 captions for explanation of pixel identification.

Figure 6: Energy resolutions for 240 MeV $^4$He particles detected by the CsI crystal. The uncorrected energy spectrum is shown in the bottom panel. It has an energy resolution of about 1.54 MeV (0.65%) FWHM. The spectrum corrected for an average gradient determined by scanning is shown in the middle panel. The energy resolution for this spectrum is 0.59% or 1.41 MeV FWHM. The spectrum corrected for local variations (pixel by pixel) is shown in the top panel. It has an energy resolution of 0.45% or 1.08 MeV FWHM.

Figure 7: Calibration curves for $^{11}$C, $^{12}$C, $^{13}$C and $^{14}$C for the CsI (Tl) crystals obtained using direct fragmentation beams listed in Table I. The curves are the best fit according to Eq. 2.

**Table I :** List of fragmentation products used in the energy calibrations of the CsI crystals.

| $^{16}$O fragmentation products | E (MeV) | $^{36}$Ar fragmentation products | E (MeV) |
|---|---|---|---|
| P | 77.17 | | |
| D | 39.78 | d | 79.57 |
| T | 26.72 | t | 53.75 |
| $^3$He | 105.0 | $^3$He | 210.00 |
| $^4$He | 79.99 | $^4$He | 160.00 |
| $^6$He | 53.64 | $^6$He | 107.90 |
| $^6$Li | 119.90 | $^6$Li | 240.00 |
| $^7$Li | 103.10 | $^7$Li | 206.80 |
| $^8$Li | 90.40 | $^8$Li | 181.60 |
| $^7$Be | 182.20 | $^7$Be | 363.40 |
| $^9$Be | 142.50 | $^9$Be | 285.60 |
| $^{10}$Be | 128.40 | $^{10}$Be | 257.90 |
| $^{10}$B | 199.90 | $^{10}$B | 400.00 |
| $^{11}$B | 182.10 | $^{11}$B | 364.90 |
| | | $^{12}$B | 335.40 |
| $^{11}$C | 261.20 | $^{11}$C | 521.60 |
| $^{12}$C | 239.90 | $^{12}$C | 480.00 |
| $^{13}$C | 221.80 | $^{13}$C | 444.40 |
| | | $^{14}$C | 413.7 |
| $^{14}$N | 279.90 | $^{14}$N | 560.00 |
| | | $^{15}$N | 524.00 |
| | | $^{16}$N | 492.40 |
| $^{15}$O | 340.80 | $^{15}$O | 680.70 |
| | | $^{16}$O | 640.00 |

|  |  | $^{17}O$ | 603.70 |
|  |  | $^{18}O$ | 571.30 |

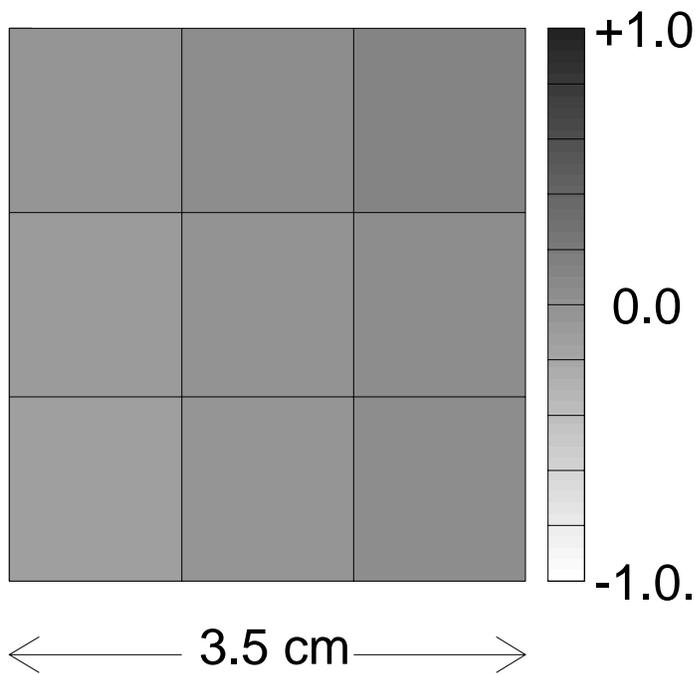

| -0.03 | +0.03 | +0.14 |
| --- | --- | --- |
| -0.09 | -0.01 | +0.05 |
| -0.11 | -0.04 | +0.05 |

Crystal 652

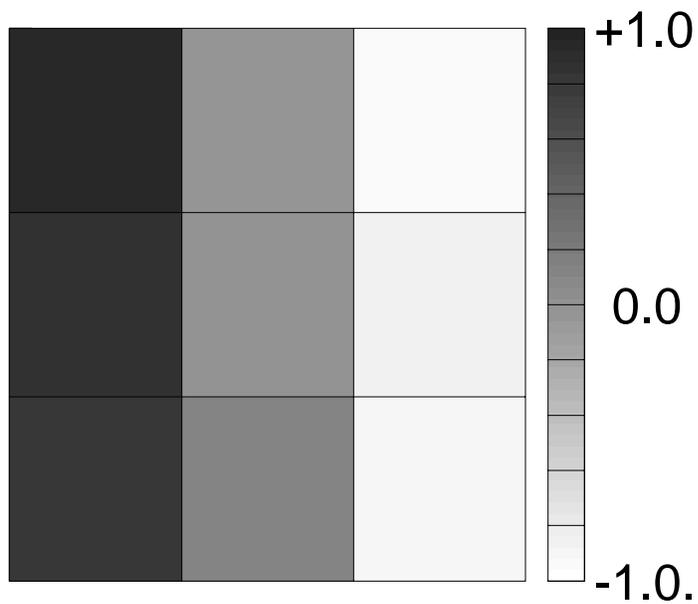

| +0.95 | -0.05 | -0.92 |
| --- | --- | --- |
| +0.88 | -0.01 | -0.87 |
| +0.80 | +0.14 | -0.91 |

Crystal 291

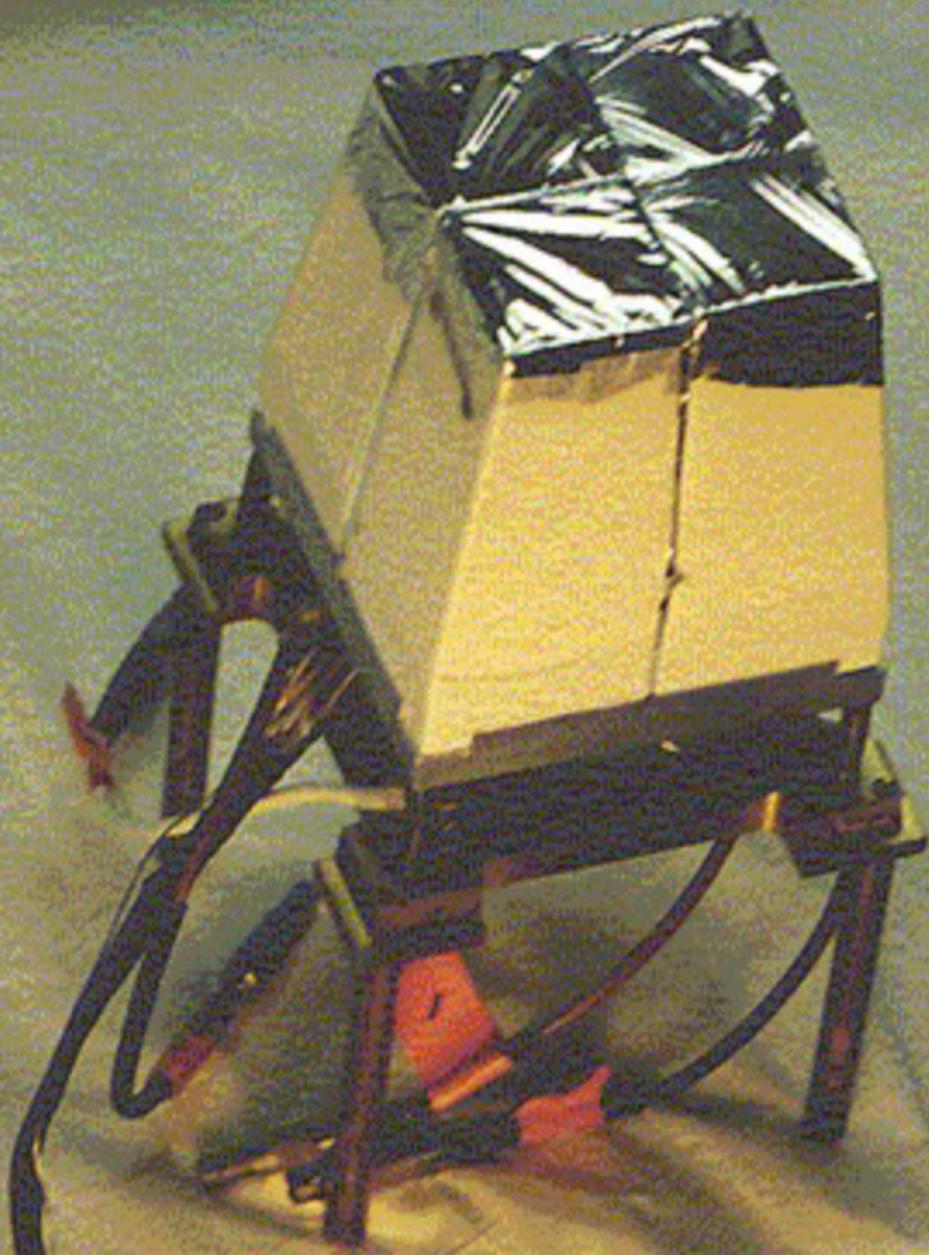

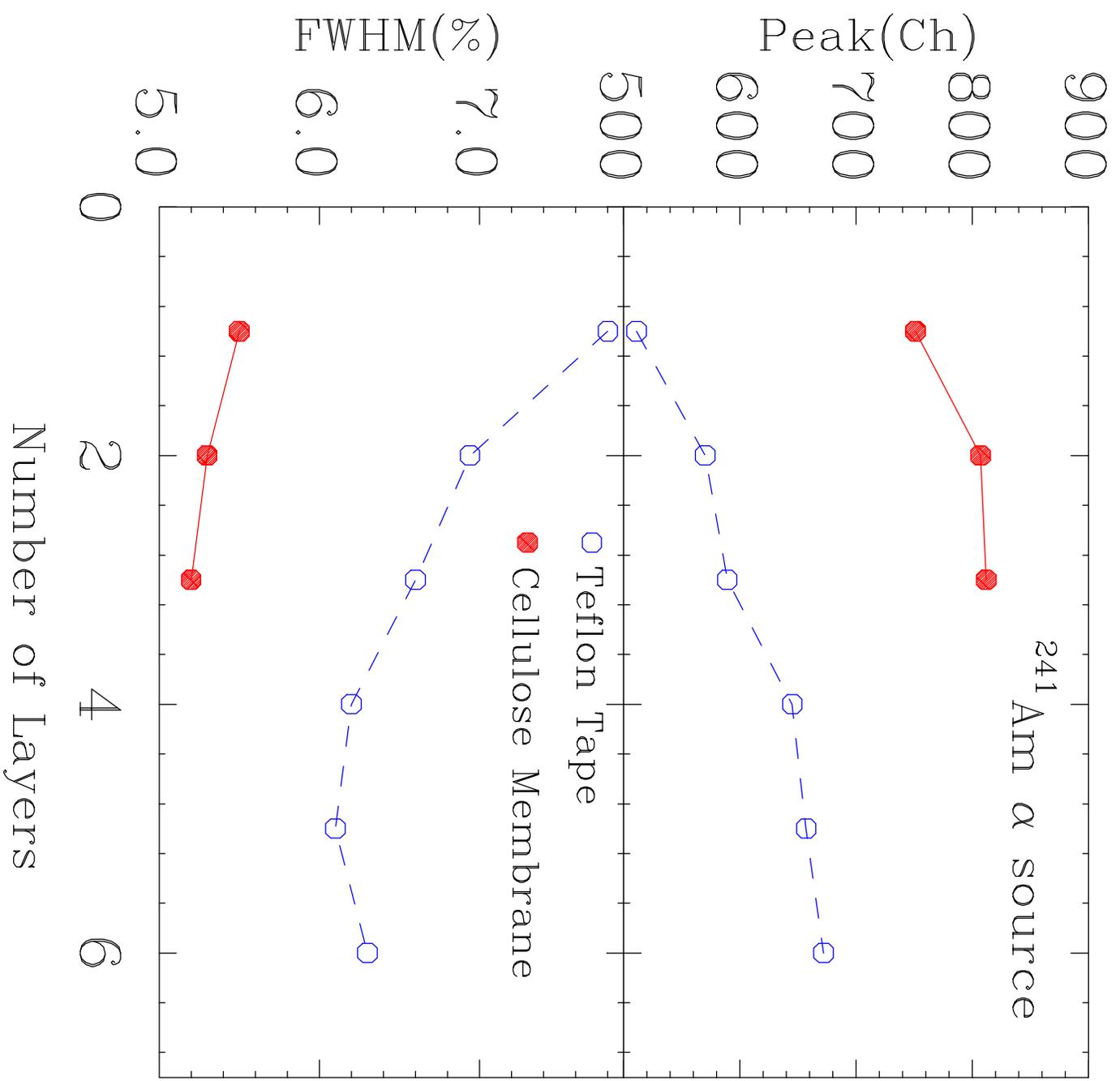

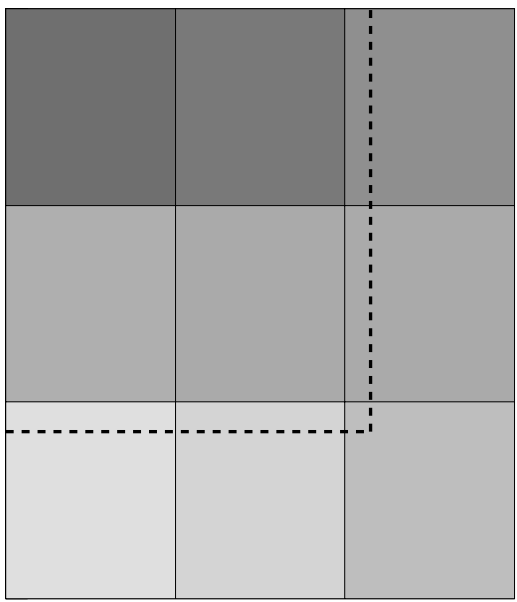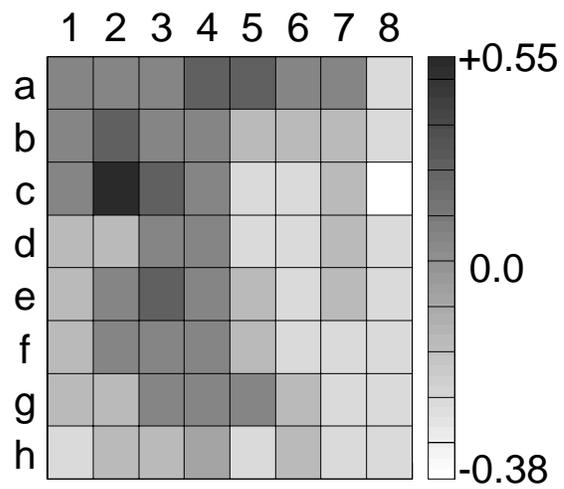

Crystal 655

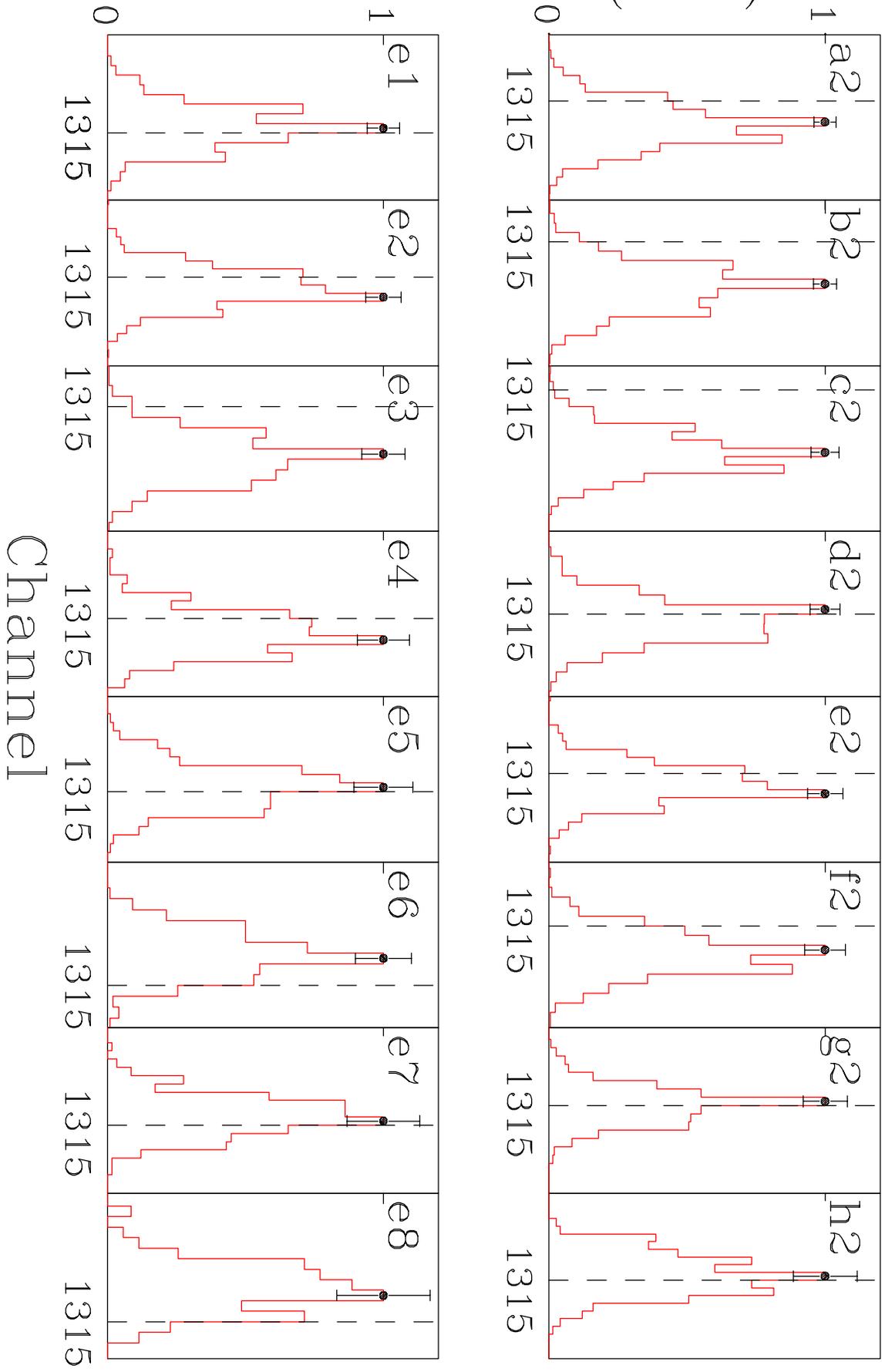

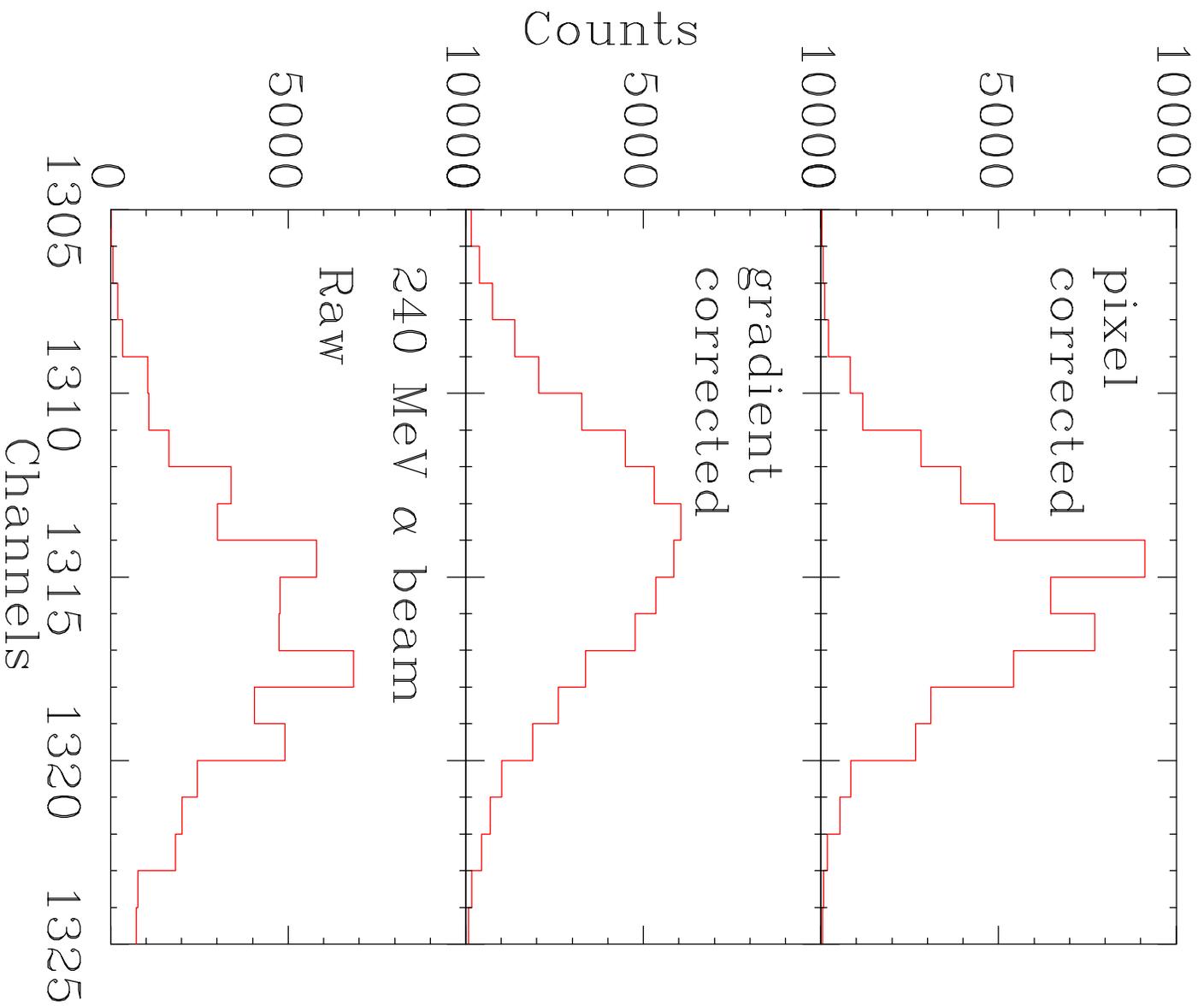

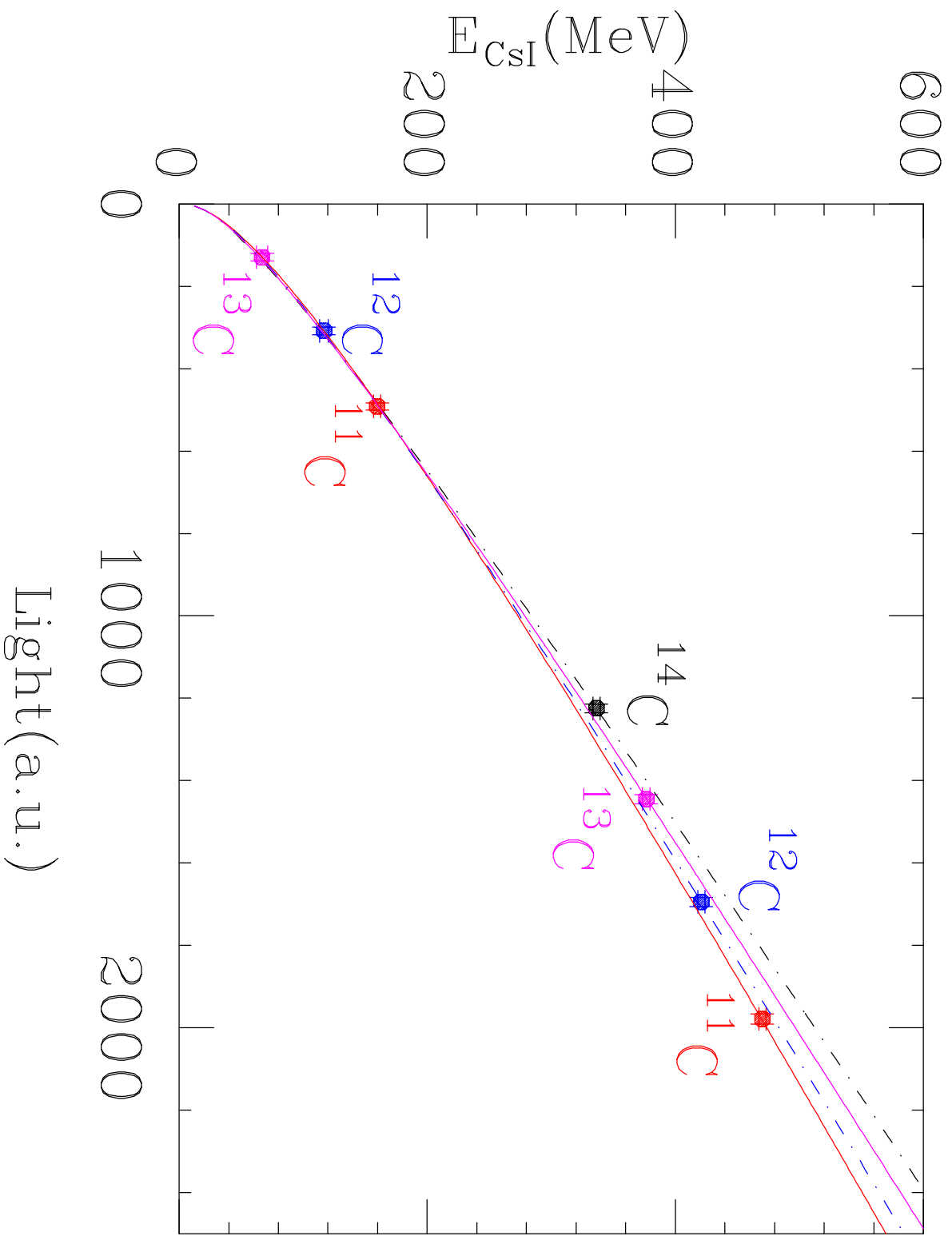